\documentclass{article}
\usepackage{natbib,latexsym,epsfig,emulateapj,onecolfloat,graphics}
\def\eps@scaling{.95}
\def\epsscale#1{\gdef\eps@scaling{#1}}

\def\plotone#1{\centering \leavevmode
    \epsfxsize=\eps@scaling\columnwidth \epsfbox{#1}}

\def\kms{\ifmmode {\rm\,km\,s^{-1}}\else
    ${\rm\,km\,s^{-1}}$\fi}
\def\ms{\ifmmode {\rm\,m\,s^{-1}}\else
    ${\rm\,m\,s^{-1}}$\fi}
\def\kmsMpc{\ifmmode {\rm\,km\,s^{-1}\,Mpc^{-1}}\else
    ${\rm\,km\,s^{-1}\,Mpc^{-1}}$\fi}
\def\hkmsMpc{\ifmmode {\rm\,h^{-1}\,km\,s^{-1}\,Mpc^{-1}}\else
    ${\rm\,h^{-1}\,km\,s^{-1}\,Mpc^{-1}}$\fi}
\def\lya{{\rm Ly}$\alpha$}

\def\kpc{{\rm\,kpc}}

\def\msun{\ifmmode {\rm\,M_\odot}\else ${\rm\,M_\odot}$\fi}
\def\Msun{\ifmmode {\rm\,M_\odot}\else ${\rm\,M_\odot}$\fi}
\def\lsun{\ifmmode {\rm\,L_\odot}\else ${\rm\,L_\odot}$\fi}
\def\Lsun{\ifmmode {\rm\,L_\odot}\else ${\rm\,L_\odot}$\fi}
\def\rsun{\ifmmode {\rm\,R_\odot}\else ${\rm\,R_\odot}$\fi}
\def\Rsun{\ifmmode {\rm\,R_\odot}\else ${\rm\,R_\odot}$\fi}

\def\cmtw{\ifmmode {\rm\,cm^{-2}}\else ${\rm\,cm^{-2}}$\fi}
\def\cmthr{\ifmmode {\rm\,cm^{-3}}\else ${\rm\,cm^{-3}}$\fi}

\def\ergps{\ifmmode {\rm\,erg\,s^{-1}}\else ${\rm\,erg\,s^{-1}}$\fi}
\def\ergpscmtw{\ifmmode {\rm\,erg\,s^{-1}\,cm^{-2}}}

\def\eg{{\it e.g.}}

\def\deg{\ifmmode {^{\circ}}\else {$^\circ$}\fi}
\def\degr{\ifmmode {^{\circ}}\else {$^\circ$}\fi}
\def\degs{\ifmmode {^{\circ}}\else {$^\circ$}\fi}

\def\etal{{\it et al.~}}

\def\Ho{\ifmmode {\rm\,H_\circ}\else ${\rm\,H_\circ}$\fi}
\def\hnot{\ifmmode {\rm\,H_\circ}\else ${\rm\,H_\circ}$\fi}
\def\h0{\ifmmode {\rm\,H_\circ}\else ${\rm\,H_\circ}$\fi}
\def\hnotunit{\ifmmode {\rm\,km\,s^{-1}\,Mpc^{-1}}\else
    ${\rm\,km\,s^{-1}\,Mpc^{-1}}$\fi}
\def\qnot{\ifmmode {\rm\,q_\circ}\else ${\rm q_\circ}$\fi}
\def\q0{\ifmmode {\rm\,q_\circ}\else ${\rm q_\circ}$\fi}
\def\ie{{\it i.e.}}

\def\vs{{\it versus} }

\def\arcsec{\ifmmode {^{\prime\prime}}\else $^{\prime\prime}$\fi}
\def\asec{\ifmmode {^{\prime\prime}}\else $^{\prime\prime}$\fi}
\def\arcmin{\ifmmode {^{\prime}}\else $^{\prime}$\fi}
\def\amin{\ifmmode {^{\prime}}\else $^{\prime}$\fi}

\def\h{{\rm h}}
\def\lesssim{\mathrel{\hbox{\rlap{\hbox{\lower4pt\hbox{$\sim$}}}\hbox{$<$}}}}
\def\gtrsim{\mathrel{\hbox{\rlap{\hbox{\lower4pt\hbox{$\sim$}}}\hbox{$>$}}}}
\let\la=\lesssim                        % For Springer A&A compliance...
\let\ga=\gtrsim

\def\be{\begin{equation}}
\def\ee{\end{equation}}

\lefthead{C. V. Manning}
\righthead{Void Matter Density}

\begin{document}
\twocolumn [

\title{The Nature and Origin of the Non-Void Ly$\alpha$ Cloud Population} 
\author{Curtis V. Manning}
\affil{Astronomy Department, University of California, Berkeley, CA
94720} 
\authoremail{cmanning@astro.berkeley.edu}

\begin{abstract}

%REVISE
%*****************************************************************
I continue my study of the low-redshift \lya\ cloud population.
Previous work showed how galaxy catalogs could be used to attribute
relative degrees of isolation to low-redshift \lya\ clouds found in
HST/GHRS spectra.  This enabled the separation of clouds into two
distinct populations corresponding to two distinct environments,
variously characterized as void/unshocked and non-void/shocked.  Void
clouds have a steep equivalent width distribution (\ie, many smaller
absorbers) while non-void clouds have a flat distribution.  I show
that N-body/hydro simulations of \lya\ clouds are inconsistent with
observations of the clouds as a function of their environments.
Simulations fail to predict the existence of significant numbers of
detectable void clouds, and incorrectly predict the characteristics of
non-void clouds.  Implicated in this failure is the so-called
fluctuating Gunn-Peterson Approximation, FGPA, which envisions that
\lya\ absorbers are formed in the large-scale structures of coalescing
matter.  A recent paper (Manning) has modeled the void cloud
population as sub-galactic perturbations that have expanded in
response to reionization.  It is notable that success in this modeling
was contingent upon using the more massive \emph{isothermal} halo in
place of the standard Navarro, Frenk \& White, for it was found that
gravitational restraint on evaporation of baryons is key to producing
detectable void absorbers.  In this paper I extend my modeling of
\lya\ clouds to non-void clouds using the same basic cloud model.  In
the case of voids, clouds are in a quiescent environment, while
non-void clouds are thought of as void clouds that have accreted to
the denser, turbulent intergalactic medium surrounding galaxies, and
so are subjected to bow shock stripping.  Model void clouds are
analytically shock stripped, and a column density spectrum (CDS) is
derived, based on the same halo velocity distribution function as that
used to explain the void CDS.  The non-void CDS produced by shocked
sub-galactic clouds are found to be capable of producing an excellent
fit to the observed non-void CDS without recourse to the FGPA
mechanism.

\end{abstract}

\keywords{intergalactic medium --- quasars:absorption lines -- dark matter}

%\clearpage
%\tableofcontents
%\newpage

]

\section{Introduction} \label{sec-intro}
N-body simulations now occupy a prominent position in the field of
astrophysics.  Their results promulgate a picture of structure
formation with which it is hard to argue.  The reason for this is
that, through a variety of means, simulations have been designed to be
consistent with high-redshift \lya\ forest spectra, and they are
integrated into the body of what can only be termed the standard
model.  However, because of the way cloud simulations are compounded,
the results at high redshift can hardly be cited as proof of the
accuracy of their simulations -- only of their self-consistency.

Recent simulations of \lya\ clouds by \citet{Dave:01} have extended
predictions of models down to a redshift of zero.  On the basis of
their simulations, they predict a column density spectrum (CDS) with a
steep cumulative spectral slope\footnote{The CDS, as well as the
equivalent width distribution function, is well-fit by a power-law
distribution.  Thus the ``slope'' is the index of the power-law.},
${\mathcal S}_{sim} = -1.15 \pm 0.04$.  Simulations generally show
that \lya\ forest absorbers arise from the mildly overdense, highly
photoionized gas that traces the dark matter potentials of large-scale
structures surrounding galaxy concentrations \citep{Bi:93,
MiraldaEscude:96, Weinberg:97, Croft:98, Riediger:98, Dave:01}.  In
this paper, I will generally refer to this as the \emph{intergalactic
medium} (hereafter IGM).  This will be contrasted with the voids,
which are not dominated by galaxies.  The mechanism for the formation
of \lya\ clouds in these simulations is the fluctuating Gunn-Peterson
approximation (hereafter FGPA), as explained in \citet{Bi:93,
Weinberg:97, Croft:98, Dave:01}.  It is based on the assumption that
gas is substantially stripped or evaporated from all smaller halos,
forming a roughly homogeneous filament surrounding the fully
non-linear coalescing structures.  These simulations have never
predicted a significant low-redshift population of \lya\ clouds in
voids; diffuse gaseous structures in voids disperse with the Hubble
flow, and today would be essentially undetectable
\citep{Riediger:98,Dave:99}.

\begin{figure}[h!] 
\centering
\epsscale{1.1} 
\vspace{-0.25in}
\plotone{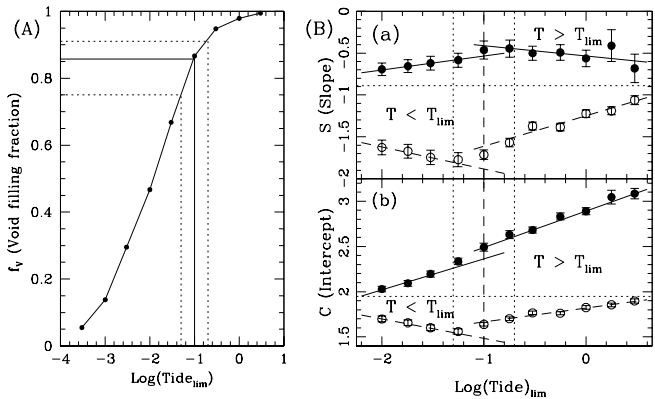}
%\plotone{/pita/manning/isolated/ewds_2upL.eps}
%%\plotone{ewds_sl_int_tidez_over.eps}
\vspace{-3.6cm}
%\plotfiddle{3b_vs_N.ps}{3.6in}{0}{50}{50}{-150}{-73} % 
\caption{\label{fig-ewdf}Panel $A$: The cumulative fraction of space
in the survey \citep{Penton:00a} with tide ${\mathcal T} \le {\mathcal
T}_{lim}$.  Assuming the sightlines are randomly selected, this is an
estimate of the volume filling fraction with tide ${\mathcal T} \le
{\mathcal T}_{lim}$.  The three vertical lines correspond to the
``transition zone'' shown in panel $B$, indicate the possible range in
value of the void filling fraction.  Panel $B$: The trend in slopes
(panel $a$) and intercepts (panel $b$) of EWDFs defined by catalogs
with tidal field ${\mathcal T} \le {\mathcal T}_{lim}$ (top of each
panel), or ${\mathcal T} \ge {\mathcal T}_{lim}$ (bottom),
respectively, for the low-$z$ cloud sample (see \S6.2.1 Paper I).  A
strong transition in the slopes in the range $-1.3 \la \log{\mathcal
T}_{lim} \la -0.7$ is apparent (two dotted vertical lines); also seen
in the intercepts. See further discussion in Paper II.}
\end{figure}

At low-redshift, galaxy redshift catalogs can be used to
assess the relative proximity of \lya\ clouds to galaxies and their
larger structures.  By so doing, it will be possible to check on the
detailed predictions of the simulations.  Therefore, the study of
low-$z$ absorbers as a function of environment may provide a crucial
test for the N-body/hydro simulations.

In \citet{Manning:02} (hereafter Paper I), low-$z$ \lya\ absorber data
\citep{Penton:00a} was used, , together with galaxy redshift catalogs,
to assess the relative isolation of low-redshift \lya\ clouds.  It was
shown how the summed scalar tidal field in space could be calculated,
so that the relative isolation of clouds could be evaluated.  I used a
limiting tidal field ${\mathcal T}_{lim}$ to divide the cloud catalog
into low, and high-tidal field catalogs.  As with Paper I, the tidal
field is given in units of inverse Hubble time squared, resulting in a
dimensionless parameter of order unity.  The equivalent width
distribution function (hereafter EWDF) of these catalogs appear to be
well-fit by power-laws.  I use the model, $\log{d{\mathcal
N}({\mathcal W})/dz}= {\mathcal C} + {\mathcal S} \log{({\mathcal
W}/63 \, {\rm m\AA})}$, where ${\mathcal N}({\mathcal W})$ is the
number of clouds per unit redshift with rest equivalent width ${\ge
\mathcal W}$.  Results of the fits can be seen in
Fig. \ref{fig-ewdf}$B$ (see caption).  It was a surprise to find that
a large population of clouds exist in extreme isolation.  Void clouds
have a steep EWDF (${\mathcal S}_{V} \approx -1.6$), while non-void
clouds have a flat slope (${\mathcal S}_{NV} \approx -0.5$). Here and
elsewhere I use the subscripts ``V'' and ``NV'' to stand for ``void''
and ``non-void''.  Thus there exist two separate populations of
clouds.  The trends of fitting parameters with variations in limiting
tide ${\mathcal T}_{lim}$ show that the two distinct types of clouds
are separated by a transition zone (vertical dotted lines), whose
width is plausibly caused by measurement errors, intrinsic scatter (in
the Tully-Fisher relation), peculiar velocities of clouds and
galaxies, and the spatial range a cloud travels during its transition
from a void-type to a non-void-type galaxy.  The dichotomy of types
that is seen in this data, especially in the difference in slopes
between void (${\mathcal T} \le {\mathcal T}_{lim}$), and non-void
EWDFs (${\mathcal T} \ge {\mathcal T}_{lim}$), is dramatic.  There
exists a similar dichotomy -- unshocked \vs\ shocked -- found in the
simulations of \citet{Riediger:98} and \citet{Cen:99}, suggesting that
the phenomenology of non-void clouds may be explained in terms of
shocks.

The cumulative distribution of tidal field strengths (${\mathcal T}
\le {\mathcal T}_{lim}$) over the lines of sight from which the cloud
data were taken \citep{Penton:00a} is shown in Fig. \ref{fig-ewdf}
panel $A$.  It is used in conjunction with the range of values of
${\mathcal T}_{lim}$ in the transition zone (Fig. \ref{fig-ewdf} panel
$B$) to assess the volume filling factor of voids.  The center of the
transition is taken to be the tidal field contour at which the
shocked andunshocked populations meet.  At $z \simeq 0$ the void filling
factor can be read off the figure;
\begin{equation} \label{eq-fvoid}
f_V = 0.86_{-0.11}^{+0.05},
\end{equation}
in substantial agreement with the model of \citet{Cen:99} ($\sim
90\%$).  This is the fraction of the universe occupied by void-type
clouds -- hence the volume containing unshocked clouds.  Void clouds
have a steep slope that , according to the analysis of
\citet{Manning:03} (hereafter Paper II), requires clouds to have flat
baryon distributions ($\rho_b \propto r^{-1.17}$), so that the
observed absorption systems are detected at relatively large impact
parameters (eg, $\langle r_p \rangle \sim 29$ kpc for $N_{HI}=10^{13}
~\cmtw$).  In the shocked regions surrounding galaxy concentrations,
such a diffuse structure would not be possible if significant cloud
motions were present.  Of course the logic behind expecting cloud
motions is that the gravitational potentials of coalescing large-scale
structure are driving them.
\begin{figure}[h!] 
\centering
\epsscale{1} 
%\vspace{-0.25in}
\plotone{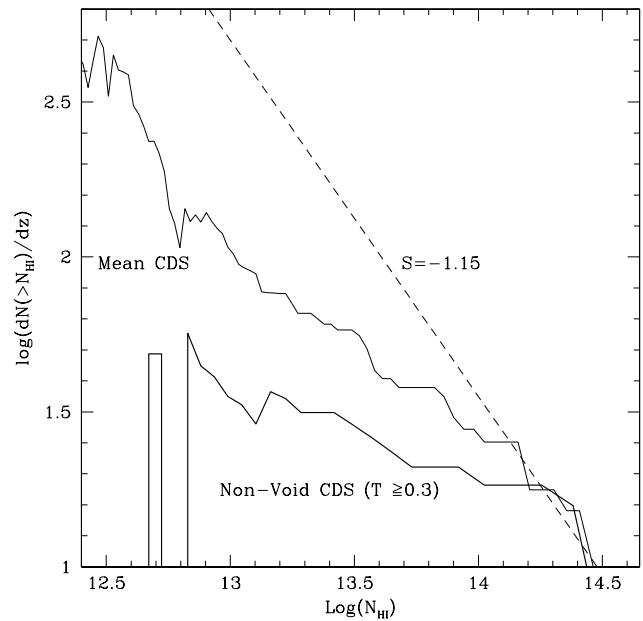}
%\plotone{colden_Dave.eps}
%\vspace{-3.6cm}
%\plotfiddle{3b_vs_N.ps}{3.6in}{0}{50}{50}{-150}{-73} % 
\caption{\label{fig-colden-Dave}The column density spectrum of the
non-void clouds (lower heavy line for ${\mathcal T} \ge 0.3$), shown
in comparison to the mean CDS (light line).  The dashed line
represents the slope of the CDS predicted by \citet{Dave:01} for
clouds that are expected to be found in regions of high tidal field.
The non-void CDS has been adjusted to represent its line density to
the CDS averaged over all space.  Over \emph{non-void} space, the
non-void CDS has a $\sim 7$ times larger line density.}
\end{figure}

Because the N-body/hydro simulations have uniformly predicted that, at
least at lower redshifts, \lya\ absorbers occupy regions in proximity
to galaxy concentrations, their predicted CDS spectral slope should
agree with the observations of the local non-void clouds, since both
are referring to clouds in the same locations\footnote{Simulations
produce a CDS, while observations produce EWDFs.  The blanketing that
occurs in clouds that are not optically thin ($N_{HI} \ga 10^{13.5}$;
Fig 9 Paper II) results in spectral slopes of CDS being generally
flatter than those of the corresponding EWDF.}.
Fig. \ref{fig-colden-Dave} is a comparison of predicted and observed
spectral slopes.  The dashed line represents the slope of the
cumulative CDS prediction of \citet{Dave:01} with an arbitrary
normalization.  Also shown are CDS derived from EWDFs according to the
method of \S5.3 of Paper II; the heavy line represents the CDS of
non-void clouds (${\mathcal T} \ge 0.3$), and the thin line is the
\emph{mean} CDS.  The predicted slope of \citet{Dave:01}, ${\mathcal
S}_{sim}=-1.15 \pm 0.04$, does not compare well to the observed
non-void EWDF slope, ${\mathcal S}_{NV}\simeq -0.5 \pm 0.1$ (Table 3
of Paper I), and agrees even less with the CDS slope $\simeq -0.36$.
Analysis shows that \emph{void} clouds are responsible for the
steepness of the mean EWDF at $z \sim 0$; low EW clouds are quite rare
in the turbulent IGM, as the cut-off in the non-void CDS in the
Fig. \ref{fig-colden-Dave} shows.  But simulations have ``detected''
low-column density clouds in the IGM where observations suggest that
their detection would be very difficult.  This suggests that the
analysis of Doppler parameters in standard N-body/hydro simulations
is incorrect.  I address this problem in more detail in
\S\ref{sec-discuss}.

The need to explain the void clouds, even while the standard model
predicts no significant numbers, drives one to re-consider the
predictions of the ``fluctuating Gunn-Peterson approximation'' (FGPA)
\citep{Bi:93,Weinberg:97,Croft:98,Dave:01}.  But \S2 of Paper II
clearly showed that the predicted Doppler parameters for clouds in
voids produced by this scenario exceed by far the observed $b$-values
of void clouds from HST/GHRS spectra \citep{Penton:00a}.  Void clouds
have Doppler parameters roughly half that of non-void clouds (Paper
I).  Clearly, this shows that void clouds cannot be expanding with the
Hubble flow.  If they are not produced by a fluctuating Gunn-Peterson
effect, then how else does one explain void clouds except as discrete
-- \ie, as subgalactic structures; remnant survivors from the
re-ionization epoch?

In Paper II void \lya\ clouds were modeled as gas associated with
sub-galactic halos that has responded to the epoch of reionization by
evaporating from their halos, albeit at a gravitationally restrained
rate.  A one-dimensional Lagrangian hydro code \citep{Thoul:95} was
used to track the evolution of the baryons following reionization at
high resolution down to a redshift $z=0$.  The products of these
simulations are used to calculate model CDS.  Two different
untruncated halo models are used, the \citet{Navarro:96, Navarro:97}
(NFW) halo, and an isothermal mass distribution ($\rho \propto
r^{-2}$).

The NFW halo is the outcome of numerical simulations, and would seem
to be the first choice for a model.  However, in the analysis of Paper
II, the NFW halo failed to restrain the evaporation of baryons
sufficiently to produce measurable column densities in a void
environment at $z=0$.  However, the isothermal halo proved viable as
long as the distribution of cloud halo circular velocities were
steep; consistent with that derived by \citet{Klypin:99, Klypin:02} for
isolated halos (see \S6.2 of Paper II).

According to N-body simulations, an isolated galaxy should have a mass
distribution in agreement with the NFW halo; a steep $r^{-3}$ density
profile for $r \gg r_{max}$.  However, this conjecture is inconsistent
with the findings that satellite galaxy distributions in groups and
clusters follow an approximately inverse square number density
relation with radius far beyond $R_{max}$ or $R_{vir}
$\citep{Seldner:77,McKay:02}, and have velocity dispersions that are
flat to similar distances \citep{Zaritsky:94, Zaritsky:97a,
Zabludoff:00, McKay:02}.  These results are indicative of an
isothermal, rather than an NFW mass distribution.

Recent re-thinking of the McKay et al. results \citep{Prada:03}
suggest the above results are affected by interloper galaxies which
are randomly distributed in space (hence in velocity or redshift).
However, that small galaxies are relatively unclustered, does not
imply that they are randomly distributed: their conjecture is in
strong contrast with findings that satellites have a number
distribution $n \propto r^{-2.1}$ \citep{McKay:02} around their
primaries.  Prada et al. claim also to have found a method to detect
interlopers placed into simulated data that can then be applied to
real data.  It is odd that though interlopers are to be randomly
inserted, their Fig. 3 shows a distinctly non-random distribution of
interlopers.  One wonders whether the algorithm for identifying
interlopers has been applied in a way to pick out things inconsistent
with the halo model.  These doubts prompt the question of whether they
merely ``discovered'' what they had presumed at the outset.  This
issue remains to be fully resolved, but for the present, I will
consider the original work (McKay et al.) to be plausible, and
probably correct in its essentials.  In support of this are two
points.  First, this assumption is at least consistent with the
results of Paper II that NFW halos were not good cloud models, and
that isothermal halos are.  Second, this particular assumption
regarding the outer regions of galactic halos is not essential to the
analysis, but it will be seen that it is consistent with other data to
be developed in \S4.3.

This same isothermal model is the basis of a successful explanation of
the void EWDF.  Perhaps non-void clouds may be explained by a similar
approach.

The goal of the current paper is to explain the nature of non-void
clouds.  I do this in terms of the environment in which they are
found.  I here describe a theory in which non-void clouds are produced
by discrete clouds.  At the epoch of their formation, sub-galactic
halos must have been distributed fairly evenly in the universe --
approximately half in mildly overdense regions, and the other half in
mildly underdense regions.  As the universe evolved, the halos in
overdense regions must be carried with the flow into the growing zones
of shocked gas, as visual presentations of structure formation
convincingly show.  Thus, according to Birkhoff's theorem
\citep{Birkhoff:23}, underdense regions came to have a locally higher
Hubble constant, which tends to promote the deepening of the
underdensity, dispersing their halos and suppressing hierarchical
growth.  Of the halos that are now in non-void space, many may have
arrived during the primary and secondary infall of the coalescing
large-scale structure.  At later times, when voids are
well-established, the ``gravitational-repulsion'' \citep{Piran:97} of
the contents of voids requires that void clouds will be ejected at
velocities dependent on void sizes and the values of the expansion
parameter in the void.

Thus, the above reasoning strongly suggests that the great majority of
current non-void clouds were, at one time or another, standard unshocked void
clouds.  Hence non-void clouds can be characterized with roughly the
same halo velocity distribution function (HVDF) as derived for void
clouds (Paper II), though the normalization will be different (see
below).

The effect of the accretion of delicate void clouds to the shocked
non-void space must be dramatic.  When the cloud encounters the
dissipated gas in non-void space, the tenuously held gas is shock
stripped, truncating the cloud, resulting in a modification of its
EWDF.  It is these effects that I intend to follow in this paper.  The
goal is to see if void clouds (\ie, as sub-galactic structures),
transformed by plausible shocks in the intergalactic medium, can be
used to explain the non-void EWDF at low-redshift.

The cosmology assumed in this paper is that of Papers I and II -- a
standard flat lambda model with $\h=0.75$.  The total matter density
of voids is either referred to as $\Omega_V={\overline
\rho}_V/\rho_{crit}$, or as $\Omega_V/\Omega_m$, where $\Omega_m =
0.3$ is assumed.  As in the other papers in this series, I assume the
cosmic baryon density is $\Omega_b/\Omega_m = 0.1$.  Recent results
from the Wilkinson Microwave Anisotropy Probe (WMAP)
\citep{Spergel:03} suggest a substantially larger value
$\Omega_b/\Omega_m = 0.166$.  The effects of the larger value will be
noted.  The large discovered line density and small Doppler parameters
of void clouds (Papers I and II) strongly suggest that they are to a
significant degree self-gravitating and discrete.  I treat them as
such herein.

I begin the analysis in \S\ref{sec-methods} by studying the effects of
shocks on model void clouds as a function of their velocity relative
to the IGM.  The non-void column density spectrum is modeled in
\S\ref{sec-NVCDS}.  In \S\ref{sec-discuss} I discuss some of the
broader issues touched upon in this paper, and summarize my findings
in \S5.

\section{Constraining Cloud Velocity}\label{sec-methods}
I model all \lya\ absorbers as initially taking the form of void
clouds, and therefore to be consistent with the models produced in
Paper II.  I view the difference in slopes between non-void EWDFs and
void EWDFs as attributable to the relatively extreme conditions of the
IGM -- a denser, hotter, and turbulent environment.  I propose that
the primary factor in the modification of cloud characteristics is ram
pressure from the relative motions of clouds through the IGM.

To constrain the rate of motion in the IGM to which clouds are
subjected, I model the effects of shocks produced by clouds moving at
a range of velocities within a medium of density $\rho_{NV}$.  What
ram pressure will explain the change in shape and relative
normalization of the CDS?  As noted above, this methodology assumes
that the transition from void to non-void cloud characteristics is
currently brought about by the impact of the clouds with the
dissipated gas near the ${\mathcal T}=0.1$ contour.

%Below I study the effects of shocks on void clouds, and see how the
%void EWDF is transformed as a function of cloud velocity..

%\newpage
\subsection{Shock Stripping of Void Clouds}  \label{sec-strip}

\citet{Murakami:94} (hereafter MI94) probed the effects of the
stripping action of blast waves on mini-halos.  Mini-halos
\citep[\eg,][]{Rees:86} were once utilized to explain the \lya\ forest
\citep{Rees:86,Murakami:93}, and are closely related to those
sub-galactic halos studied in the present work.  MI94 envisioned two
phases of stripping -- the first being the immediate result of the
onset of the shock, and the second being the gradual stripping of the
remnant by the relative velocity through the medium.  MI94 found that
the isothermal mini-halos suffer significant erosion during the
initial shock only where the internal pressure of the cloud is less
than the ram pressure.  For the case of a cloud with velocity
$v_{cl}$, encountering a dissipated medium with gas of density
$\rho_{g}$ the ram pressure on the cloud is,
\begin{equation} \label{eq-pram}
p_{ram} \equiv \rho_{g} v_{cl}^2.
\end{equation}
MI94 showed that mini-halos could withstand the secondary stripping
for a sustained amount of time if the escape velocity from the cloud
were in excess of the oblique flow within the shock.  Some clouds were
seen to survive in excess of 3 Gyr.

\begin{figure}[h!] 
\centering
\epsscale{1} 
\plotone{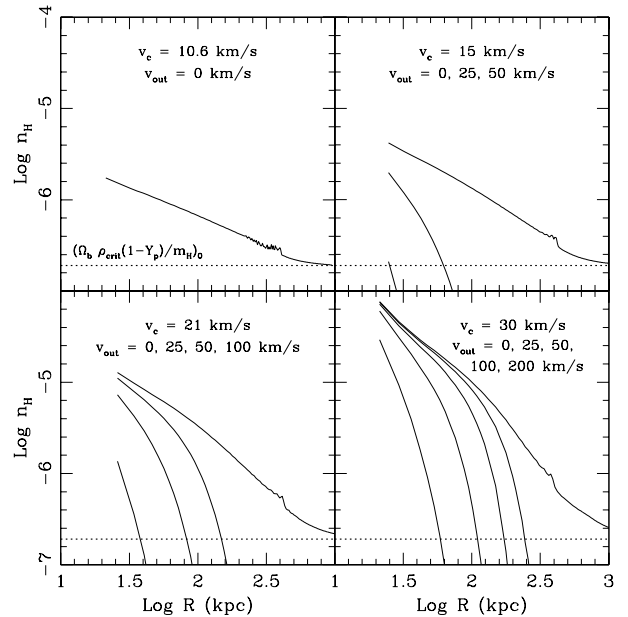}
%\plotone{denisogx_cmpr.eps}
\caption{\label{fig-cmpr-strip} The density profiles of grown halos
processed according to the prescription of Eq. \ref{eq-stripping} for
various velocities $v_{cl}$ (as noted in the panels).  The dotted line
represents the hydrogen density in the background (critical density)
medium.  When the cloud halo circular velocity $v_c = 10.6 $ \kms, the
cloud is totally stripped even at $v_{cl}=25 $ \kms.  Larger clouds
(higher $v_c$) survive larger ram pressures.  In the final panel,
cross-section of the $v_c=30 $ \kms\ cloud when $v_{cl} = 200 $ \kms\
is only about 1/4 of that when $v_{cl}=100$ \kms. }
\end{figure}

The initial cloud models are produced according to the methods
explained in \S6 of Paper II.  In these simulations, 200 baryonic bins
are used.  The first strip, which removes gas at pressures less than
the ram pressure, can be implemented by the following transformation
of density in each Lagrangian bin,
\begin{equation} \label{eq-stripping}
\rho(i) \Longrightarrow \rho(i	) \, e^{-(p_{ram}/p(i))},
\end{equation}
where the pressure in the cloud in the $i$th bin is $p(i)$.  This
formulation produces a fairly abrupt truncation of the baryon density
where the cloud pressure is less than the ram pressure.  It is assumed
that the density of gas in the non-void
medium (see Eq. \ref{eq-pram}) is,
\begin{equation} \label{eq-rhogf}
\rho_{NV} \approx 2 ~\frac{\Omega_b}{\Omega_m}~\rho_{crit},
\end{equation}
where $\rho_{crit}=3H^2/8 \pi G$, a function of redshift.  This
value is in recognition of the fact that this region is undergoing a
mildly non-linear collapse, and therefore the average density should
be mildly super-critical -- ie, $\sim 2 ~\rho_{crit}$.

In Fig. \ref{fig-cmpr-strip} I show the results of subjecting
un-stripped model clouds (top line in each panel) to ram pressures of
a gas of the above density, and outfall velocities 25, 50, 100, and
200 \kms.

\subsection{Bow Shock Stripping}  \label{sec-bow-shock}
When the cloud outfall velocity is greater than the sound speed of the
ambient medium, a bow shock will result.  I assume the IGM has an
average temperature $T_{NV}=10^4$ K, consistent with an adiabatic
sound speed of $c_s \sim 17$ \kms.  This temperature is consistent
with the simulation of \citet{Dave:01} for an over-density $\sim 4 \,
\rho_{crit}$.  In sustaining the first shock, the cloud has been
stripped down to a volume within which the pressure is greater than
the ram pressure of the shock.  From this point, a quasi-stable
situation results; a bow shock is set up which stands off some
distance from the cloud proper.  The ambient gas penetrates the shock,
and is compressed and heated, according to the Rankine-Hugoniot jump
conditions, producing a shear layer that meets with the cooler, denser
medium of the cloud.  There is a high-pressure zone at the head of the
cloud due to the presence of the moving cloud immediately behind it,
causing gas to be diverted around it.  Gas is then accelerated down
the flanks of the cloud by the pressure differential.  The transverse
motion over the cooler and denser cloud body may cause
Kelvin-Helmholtz (K-H) instabilities which could strip away the cloud,
layer by layer.

For plane shocks in the Mach range of $3 \le {\mathcal M} \le 6$
($\sim 50$ to 100 \kms), the velocity inside a plane shock has Mach
numbers ranging from 0.49 to $\sim 0.46$, respectively.  Oblique
shocks produce higher Mach values inside the shock.  This fact, plus
the pressure gradient along the length of the cloud inside the bow
shock means that the velocity transverse to the cloud may
substantially exceed Mach 0.5, perhaps by a factor of 1.5 or so.  At
the same time, the contrast between the average cloud density in the
un-stripped part of the cloud and the gas between the cloud and the
bow-shock varies from $\sim 50$ to 64 to 90, for $v_{cl}=50$, 75, and
100 \kms, respectively.  According to the 1-D analysis of
\citet{Vietri:97}, the growth rate of K-H perturbations at $0.7 \le
{\mathcal M} < 1.0$ is essentially zero for adiabatic fluids when the
density contrast between the cloud and the stripping medium is on
order 100.  These considerations suggest that stripping rates are low
for the outfall velocities we are considering.

Anecdotal evidence also suggests that clouds may endure the stripping
of the diffuse IGM gas for long periods without significant mass
erosion.  For instance, the very existence of the Local Group cloud
compact high velocity cloud CHVC 125+41-207 at a distance $50 \la d
\la 137$ kpc \citep{Bruns:01}, with its dense core and a long tail,
strongly suggests it has survived billions of years of stripping.
For, since CHVCs are found to have an average velocity $\sim -100 $
\kms\ relative to the barycenter of the Local Group \citep{Braun:99},
this cloud would have to have existed for up to $\sim 10$ Gyr in order
to have moved a Local Group diameter of $\sim 1$ Mpc.  Since there is
no conceivable way of creating this cloud within the Local Group in
recent times, we must conclude that it has been traveling through the
IGM at a similar velocity for a time on order 10 Gyr, retaining much
of its mass.  Thus this and other CHVCs may be ancient objects that
have survived billions of years of bow shock stripping; they could be
the dissipatively stripped and compacted ancestors of secondary
infall, as explained in the scenario of \citet{Manning:99a}.

While it is reasonable to postulate that smaller clouds might not
endure this stripping as long as larger clouds, there is no way at
present, short of detailed hydro/gravity simulations, to precisely
determine the lifetimes of clouds under bow shock stripping.  However,
we may assume they all survive on order 10 Gyr, understanding that
this may over-estimate the quantity of surviving clouds in non-void
space at the present time.  However, since clouds with low halo
velocities are stripped much more efficiently by shocks, large clouds
dominate the cross-section of absorbers at the velocities relevant in
this problem, so that small clouds will have only a small effect on
results in any case.

\begin{figure}[h!] 
\centering
\epsscale{1} 
%\plotone{/pita/manning/1dcode/currentcode/grow_lagr/denfls2/see below.
%\plotone{denisogxth200.eps}
\plotone{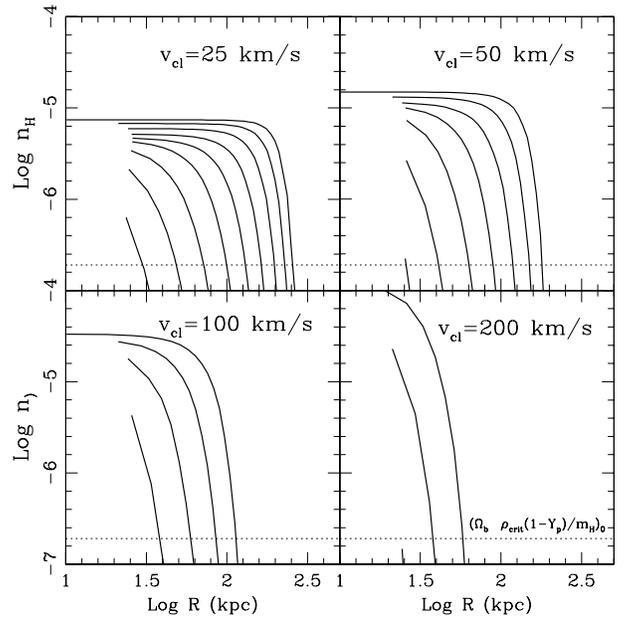}
\caption{\label{fig-cmpr-flatstrip} The density profiles resulting
from the redistribution of baryons to an average density within the
zone where the cloud pressure is greater than the ram pressure, then
smoothed at the edge with same exponential used in
Eq. \ref{eq-stripping}.  Shown are the distributions for the range of
halo velocities noted in \S\ref{sec-turb}, in ascending order, left to
right.  These profiles provide the fiducial cloud model for the
calculation of the non-void CDS.}
\end{figure}
\newpage
\subsection{Turbulent Mixing of the Surviving Cloud}\label{sec-turb}

In the globular cluster formation model of \citet{Manning:99a}, the
internal baryonic density of self-gravitating clouds that are stably
responding to supersonic winds (\ie, clouds are large enough to
survive both stages of stripping) must increase with time.  This is
partly due to the deceleration of the cloud by ram pressure, but is
also the result of a cycle of pressure-heating at the front-end of the
cloud, and radiative cooling, as the denser cloud gas is pushed along
toward the trailing end of the cloud by the shear wind.  The K-H
instabilities may gradually introduce an orderly, toroidal convection
pattern in the cloud.  This physical picture may be very similar for
void clouds entering the non-void environment.  The effect of both
cloud deceleration (including a displacement of the baryonic cloud
from its dark halo), and the fluid waves driven down the ``fetch'' of
the cloud's edge, is to disrupt the previous density profile within
the truncation radius of the shock.  In the simulations of MI94, the
central condensation of the baryons within the bow shock is quickly
lost.  There are thus good grounds for assuming that the baryons
within the shock (excepting the shear layer in between) become more
evenly distributed; cloud containment switches from gravitational to
pressure confinement when moving into non-void space.

For a large cloud, a velocity of 100 \kms\ will truncate the cloud at
about 100 kpc (see Fig. \ref{fig-cmpr-flatstrip}).  I use these values
as fiducial in the calculation of the time-scale for the
transformation from a void cloud to a non-void cloud.  A useful
conversion for the velocity is $100 ~\kms \sim 100$ kpc/Gyr.  Thus, it
would take $\sim$2 Gyr for the cloud to pass completely into the
dissipated gas of the filaments.  During, and after the process of
shock stripping, the baryonic cloud is being decelerated by the ram
pressure.  I calculate the time required for the baryonic cloud to be
slowed enough for it to become displaced from the DM by one cloud
radius.  DM plays an important role in maintaining the central
condensation of the cloud, and when it is gone, pressure gradients can
quickly redistribute the gas.  The deceleration caused by ram pressure
is,
\begin{equation} 
a = \frac{\rho_{NV} \pi (r_{cl} v_{cl})^2}{m_{cl}}.
\end{equation}

I am interested in when the displacement is one cloud radius; $r_{cl}
= 0.5\, a t^2$.  Solving,
\begin{equation} 
t = \left(\frac{r_{cl}}{v_{cl}}\right) \,\sqrt{\frac{8 \rho_{cl}}{3
\rho_{NV}}}.
\end{equation} 

Fiducial values for the density of dissipated gas in the IGM are,
$\rho_{NV} = 2 \Omega_b \rho_{crit}$, and $\rho_{cl} \approx 10
\Omega_b \rho_{crit}$.  Thus, since $r_{cl}/v_{cl} = 1$ Gyr, $t
\approx 3.6$ Gyr.  In \S\ref{sec-discuss}, I show that the
characteristic half-width of the transition zone is $\sim 0.5$ Mpc
(for $0.1 \le {\cal T} \le 0.2$).  For average cloud baryon densities
${\overline \rho}_{cl} \la 20 ~\Omega_b \rho_{crit}$, the transition
time-scale $t \la 5$ Gyr.  The cloud thus moves less than the
half-width of the transition zone by the time it is effectively
flattened.

This general picture is confirmed by Fig. 2 of MI94; by the time of
Figure 2 $d$, $\sim 1.7$ Gyr after the initial shock, a shear layer
envelops a cloud with a flat density profile, bounded by
sharp density gradients.  The loss of the central condensation of this
model cloud was accomplished in about one half the time estimated above.

It is thus expected that the clouds in ``undisputed'' non-void space
(ie, $\log{{\cal T}} \ga -0.7$; outside the transition zone) will
approximate a flattened, or random mass distribution inside the shear
layer of the cloud, so that the average density in random sightlines
is independent of the cloud impact parameter.  In the transition
zone, one may find void clouds, clouds in various stages of the
process of being transformed into non-void clouds.

To account for this effect in model clouds, I re-distribute the mass
within the truncated cloud so that the density is uniform, but I apply
the same exponential factor used in Eq. \ref{eq-stripping} to smooth
the edges.  In addition, the neutral fraction of hydrogen in the cloud
must be ``flattened'' as well; I substitute the mass-weighted neutral
fraction within the truncated cloud summed over each bin in the
truncated cloud.  Figure \ref{fig-cmpr-flatstrip} shows the density
profiles derived using this methodology for various outfall velocities
$v_{cl}$.  The halo velocities corresponding to the various lines in
the figure are given by
\begin{displaymath}
v_c = 5.31 \times 10^{0.05 \, n} ~\kms,
\end{displaymath}
where $n = 16$ (\ie, $33.5 $ \kms) for the far right-hand side
line\footnote{in the modeling of Paper II, this is the largest halo to
survive without inside-out collapse, and star formation}, $n=15$ for
the next at 29.9 \kms, then 26.6, 23.7, 21.1 \kms,.., and so on, for
$n=14$, 13, 12, etc.

%[EDITOR, PLACE FIG. 5 HERE]
\begin{figure}[h!] 
\centering
\epsscale{1.5} 
\plotone{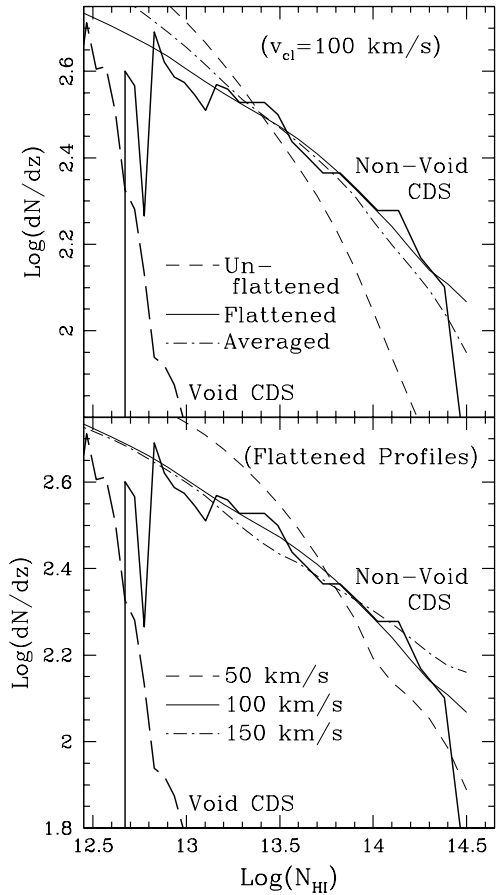}
%\plotone{/pita/manning/1dcode/currentcode/grow_lagr/fig5_2.eps}
\caption{\label{fig-cdspectgthm} Normalized model column density
spectra in relation to the observed non-void CDS (heavy solid line)
and void CDS (heavy long-dashed line).  The upper panel presents model
CDS for unflattened halos (short-dash; profiles as in Fig. 3),
flattened halos (thin solid line; profiles as in Fig. 4), and that
resulting from a profile made by averaging the two (dot-dashed line).
In this figure, the cloud velocity is $v_{cl}=100$ \kms.  In the lower
panel, model CDS are presented for various cloud velocities using
flattened cloud profiles (as in Fig. 4).  For $200 \ga v_{cl} \ga
75$\kms, these cloud profiles provide a good fit to the slope of the
observed non-void CDS.  The normalization factors $f_{mult}$ are shown
in Fig. \ref{fig-vout-fmult}.}
\end{figure}

\section{Modeling the Non-Void CDS} \label{sec-NVCDS}

The CDS is produced using a method very similar to that used to model
the void clouds and the HVDF to produce the void CDS (\S6, Paper II).
The difference in treatment is that there is an extra variable -- the
cloud velocity $v_{cl}$ that results in the stripping of clouds.  Each
model CDS is initially normalized using $\phi_V^*$, the void LF
normalization, and must be adjusted to approximate the observed CDS
with a ``concentration factor'' $f_{mult}$, which is a function of
cloud velocity.  

The first step in modeling the non-void CDS is deciding which density
profile of the shocked clouds works best.  The upper panel of
Fig. \ref{fig-cdspectgthm} shows normalized CDS for a shocked cloud
($v_{cl}=100$ \kms) in which the density profile is unadjusted, as in
Fig. \ref{fig-cmpr-strip} (short-dashed line), one in which it is
flattened, as in Fig. \ref{fig-cmpr-flatstrip} (solid line), and an
average of the two profiles (dot-dashed).  This average is produced by
simply averaging the number of H atoms in the respective Lagrangian
bins of the unflattened and flattened clouds.  Note that the range of
legitimacy of this gas profile is expected to be confined to the
transition zone since clouds are expected to be fully flattened by the
time they emerge from it.  These CDS are shown in relation to the
observed non-void CDS (${\cal T}_{lim} \ge 0.1$ (heavier jagged line).
For reference, the void CDS is represented by the long-dashed line.
The quality of the fit to the CDS is consistently good over the range
$75 \la v_{cl} \la 200$ \kms.

These results suggest that the flattened ram-pressure stripped cloud
profiles described above produce the best fit to the non-void CDS.
Thus, results confirm expectations.  I therefore adopt the flattened
profiles as the preferred non-void cloud model, realizing it may not
work well in the transition zone.

The next step is to constrain $v_{cl}$.  The lower panel of
Fig. \ref{fig-cdspectgthm} shows normalized model CDS multiplied by
the adjustable factor $f_{mult}$ based on cloud velocities of 50, 100,
and 150 \kms, shown in comparison with the observed non-void CDS, as
derived in Paper I.  Note that the high quality of the fit to the
observations shown when $v_{cl}=100$ or 150 \kms, is absent for a
velocity as low as 50 \kms.  

\begin{figure}[h!] 
\centering
\epsscale{1} 
\plotone{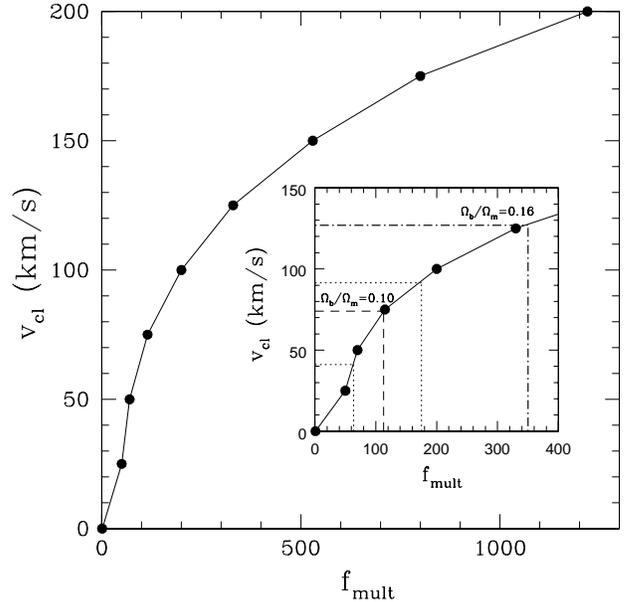}
%\plotone{vout_fmult.eps}
\caption{\label{fig-vout-fmult}The relation between the cloud velocity
$v_{cl}$, and the multiplicative factor $f_{mult}$ necessary to fit
the model CDS (made by simulating the stripping process) to that of
the observed.  The significance of the lines are explained in the
text.}
\end{figure}

Figure \ref{fig-vout-fmult} shows the concentration factors $f_{mult}$
required to adjust model CDS, with their void HVDF normalization, to
match the observed non-void CDS for a range of cloud velocities.  Such
a concentration factor is the natural result of models of hierarchical
clustering (\citealp[e.g.,][]{Lacey:93}).

%[EDITOR, PLACE FIG. 6 HERE]

It is possible to estimate the value of the concentration factor
independently from the relative values of the normalizations of the
HVDFs in void and non-void space.  According to \S7 of Paper II, the
void CDS can be best explained with the HVDF of the ``grown'' halo
cloud profiles, with slope parameter $\alpha \simeq -1.95$, and with
normalization $\phi_V^* \simeq 0.06 ~\phi^*$ (\S6, Paper II), where
$\phi^*$ is the normalization of the mean luminosity function (LF).
With the $B$-band LF, I use $\phi^*\simeq 0.022 ~\h^3 \,{\rm Mpc}^{-3}
= 0.0093 ~\h_{75}^3 \, {\rm Mpc}^{-3}$ (Eq. 36 of Paper II).  Knowing
the filling factors, we can write the equation,
\begin{equation} \label{eq-phi-ff}
\phi_V^* f_V + \phi_{NV}^* f_{NV} = \phi^*,
\end{equation}
where $\phi_{NV}^*$ is the normalization of the LF in the non-void
space (again functionally identical to that of the region
containing shocked gas), with filling fraction $f_{NV}$, and $f_V$ is
given by Eq. \ref{eq-fvoid} (\ie, $f_{NV} = 1-f_V$).  Thus,
\begin{equation} \label{eq-phiF}
\phi_{NV}^* = \phi^* \frac{1- f_V \,(\phi_V^*/\phi^*)}{f_{NV}}=
6.77_{-2.95}^{+3.73} ~\phi^*,
\end{equation}  
where the errors are propagated from the range in $f_V$ in
Eq. \ref{eq-fvoid}.  Using the void normalization $\phi_{V}^*$ for the
grown halos, the expected concentration factor required to arrive at
the normalization of the non-void population (Eq. \ref{eq-phiF}) would
be
\begin{equation} \label{eq-fmult}
f_{mult} = \frac{\phi_{NV}^*}{\phi_V^*} \approx 113_{-49}^{+62},
\end{equation}
though this is subject to a few caveats (see below).

At this point it is convenient to consider how things would appear if
I had used $\Omega_b/\Omega_m=0.16$, consistent with the WMAP data
\citep{Spergel:03}, instead of 0.10 (\S\ref{sec-intro}).  I re-ran the
simulations discussed in Paper II using this new value, and derived
the CDS, which is consistent with the observed void CDS.  The fit was
managed using the same faint-end slope parameter $\alpha=-1.95$, but
the normalization was lower; $\phi_V^*\simeq0.02 \, \phi^*$, $\sim
1/3$ of the value derived with the lower baryon density.  These model
clouds were analytically shock-stripped as described above, resulting
in very similar values of $f_{mult}$ for a given $v_{cl}$; apparently
the larger cloud neutral fraction, and consequently larger
cross-section at a given column density, make up for their lower
indicative space density in voids when $\Omega_b$ is larger.  However,
because $\phi_V^*$ is now lower, Eqs. \ref{eq-phiF} and \ref{eq-fmult}
have different values.  The non-void normalization is slightly higher
($\phi_{NV}^* = 7.02_{-3.08}^{+3.88} \, \phi^*$) and the predicted
concentration parameter is about 3 times higher ($f_{mult} \simeq
351_{-154}^{+194}$).

Due to the transferral of matter from void to non-void space, the
distributions of halo velocities of void and non-void clouds are
causally linked.  If the slope of the HVDF does not change going from
the unshocked to the shocked environments, and if void clouds are
indeed the precursors of non-void clouds, then the multiplicative
factor needed to adjust the CDS of stripped clouds to the observed
non-void CDS is given by Eq. \ref{eq-fmult}.

However, Eq. \ref{eq-fmult} should not be accepted uncritically.
\citet{Klypin:02} found that sub-halos have HVDFs consistent with a
slope parameter $\alpha\simeq -1.72$, while isolated halos have
$\alpha \simeq -1.95$ (see \S6.2 of Paper II).  This is suggestive of
a flattening of the spectral slope in higher-density areas.  If the
observed non-void clouds have average halo circular velocities of $v_c
\simeq 30$ \kms, then for $\alpha=-1.72$, there would be 2.7 times
fewer 30 \kms\ clouds in non-void space than if $\alpha=-1.95$.  This,
in turn, would require a reduction of $f_{mult}$ by a factor $\sim
2.7$.  That is, Eq. \ref{eq-fmult} should, by rights, have been
$f_{mult} = \phi_{NV} (v_c\sim 30)/\phi_V(v_c \sim 30)$, for the
greater effect of stripping on small clouds results in the largest
clouds providing the bulk of the cross-section of stripped clouds.
However, we have no direct information on their number density in the
IGM, as we do of ${\mathcal L}^*$ galaxies.  Though the higher
densities of clouds in non-void space gives greater opportunity for
hierarchical agglomeration and a consequent flattening of the spectral
slope, these clouds are by no means generally sub-halos, for the
overwhelming majority of non-void clouds in this survey are far
outside the virial radius of any galaxy (see \S\ref{sec-discuss}).

On the other hand, it is more plausible that ${\mathcal L}^*$
galaxies, lying in dense zones of the filamentary structures, could
have enhanced numbers (relative to the initial, but concentrated halo
spectrum) due to hierarchical agglomeration of smaller halos --
especially of sub-perturbations accreted during the primary collapse
phase at $z \ga 3$.  I consider it plausible, therefore, that the
$f_{mult}$ applicable to clouds may be up to $\sim 3$ times smaller
than quoted in Eq. \ref{eq-fmult}.

Fig. \ref{fig-vout-fmult} shows the trend of $f_{mult}$ with $v_{cl}$
neglecting this possible factor.  Over the range of velocities of
relevance to this study, the concentration factors for the larger
baryon density ($\Omega_b/\Omega_m \sim 0.16$) are entirely consistent
with those of the smaller baryon density.  The inset shows a set of
three vertical lines corresponding to the values of $f_{mult}$ in
Eq. \ref{eq-fmult}, and imply cloud velocities in the range $42 \la
v_{cl} \la 91$ \kms.  The dot-dashed line represents the central value
of $f_{mult}$ when $\Omega_b/\Omega_m=0.16$, implying a central value
$v_{cl} \approx 130$ \kms, and a range $100 \la v_{cl} \la 150$ \kms.

As noted above, the considerations of hierarchical merging in dense
zones suggest that $f_{mult}$, as calculated by Eq. \ref{eq-fmult},
may be less than indicated by Eq. 7 (\ie, 113 and 350 \kms\ for
$\Omega_b/\Omega_m=0.10$ and 0.16, respectively), perhaps by a factor
on order 3, although the relative isolation of most non-void clouds
suggest it is a small fraction of this.  According to
Fig. \ref{fig-vout-fmult}, this would imply a significantly smaller
velocity.  For $\Omega_b/\Omega_m=0.10$, this would imply a velocity
$v_{cl} \ll 50$ \kms.  Recall, however, that the quality of the fit of
the model non-void CDS to the observed declines when $v_{cl}\la 75
\kms$.  For the larger $\Omega_b$, cloud velocity could be expected to
drop from $\sim 130$ \kms\ to $\ga 80$ \kms, and still give a good fit
to the observed CDS slope.

The one thing that may argue against this whole logical construct is,
of course, that non-void clouds can be understood in terms of a FGPA.
I visit this issue in the next section.

\section{Discussion and Summary} \label{sec-discuss}

Further discussion in three broad areas is needed to tie this
investigation together; the FGPA, the transitionn zone, and the
line-density of absorbers as a function of ${\cal T}$ in non-void
space.

\subsection{FGPA}

It was shown in \S\ref{sec-intro} that the predictions of the FGPA for
the IGM do not appear to agree with observations.  The presence of
many clouds in void space (Paper I) stands in contrast with their
relative absence under the FGPA.  Their absorbers are produced in the
slowly varying dissipated gas of the filamentary structure surrounding
galaxy concentrations.  However, the distribution of clouds proponents
predict is inconsistent with the observed distribution of clouds in
the same non-void locale (see Fig. 2).  These indicate a significant
problem with the standard model of low-$z$ \lya\ clouds, and hence, by
extension, perhaps also with that of the high-$z$ \lya\ forest.

Although the present work has shown that non-void clouds may be
explained by the effects of shocks on previously unshocked void
clouds, this does not mean that none of the absorbers have their
origin in an FGPA.  The above analysis suggests that if they do exist,
the FGPA absorber Doppler parameters should be larger than those of
sub-galactic halos, which are essentially self-gravitating and whose
integration path lengths are smaller.  Similarly, it appears likely
that the prediction of very low column density clouds in the IGM is
due to an incorrect assessment of the Doppler parameters.  For among
the initial assumptions of the FGPA is that of neglecting turbulent
effects within the line (which is already broadened by relative
velocities of order $10$ to $40 ~\kms$), as explained in
\citet{Bi:93}.  Required integration path lengths are on order 1 Mpc
at $z\sim 0$ (Paper II).  It seems improbable that in a turbulent
medium there would not be a broadening of the spectral absorption
lines over distance such as this.  

What sources for turbulence might there be?  If the ``Birkhoff''
effect is propelling centrally condensed clouds into the IGM, this
represents a major source of kinetic energy to drive the turbulence.
On the high ${\cal T}$-side of the non-void universe, energy injection
comes with superwinds from post-starburst galaxies and the like.  If
turbulent motions of $\sim$100 \kms\ ($v_{cl}$) exists over scales of
200 kpc ($2\,r_{cl}$), then over megaparsec scales, a significant
turbulent dispersion of absorption lines on order $\ga100$ \kms\ would
be plausible.  Their Doppler parameters are in fact large enough to
make the lines appear as a continuum depression.  Since FGPA
integration path lengths are weakly correlated with column density,
FGPA Doppler parameters should be rather uniformly $\ga 100$ \kms.  By
contrast, the Doppler parameter histogram of non-void clouds finds $b
\simeq 60$ \kms\ with a dispersion of 15 \kms\ or so (Fig. 7 of Paper
I; Fig. 15$b$ of Paper II).  These histgrams show no sign that there
is a higher-$b$ component of absorbers.  In fact, the $60 ~\kms$
broadening seems appropriate for a centrally condensed cloud
undergoing a dissipative interaction with the IGM.  Thus, there is no
sign of FGPA clouds at $b \la$ 100 \kms, the upper-limit for the
\citet{Penton:00a} data.

\subsection{The Transition Zone}

Much of the modeling of non-void clouds depends on the physical
picture of a sudden transition from a cool diffuse void to a warm,
reltively dense, and turbulent IGM.  Let us consider this picture in
some detail, motivated by the following question: why should non-void
space have a relatively abrupt end, so that clouds are quickly
shock-stripped?  The fundamental fact about the edge of voids is that,
on the one hand, diffuse, adiabatically cooled matter is balanced by a
denser and warmer gas on the other.  Obviously, the pressure in the
latter $p_{NV}$ is much greater than that in the former $p_V$.  The
pressure difference will cause an expansion of the IGM into the
diffuse void gas.  Alternatively, the fact that the local expansion
parameter in voids is greater than the mean, suggests that we may
consider that the cool, diffuse gas is flowing into the denser gas at
a velocity sufficient to cause stationary shock front.  The equation
for pressure balance is then,
\begin{equation}
\rho_V v_{out}^2 = p_{NV} - p_V,
\end{equation}
where $v_{\rm out} = (H_V-H_0) R_V$.  Solving for $v_{\rm out}$ in
Eq. 10,
\begin{equation}
v_{out} = \sqrt{\frac{k}{\mu m_H} \left(\frac{\rho_{NV}
T_{NV}}{\rho_V} - T_V\right)}.
\end{equation}
The ambient temperature of non-void space is assumed to be $10^4$ K
(see \S\ref{sec-bow-shock}), while the adiabatically cooled void space
is $\sim 3000$ K (Paper II, \S4.3 and Fig. 5).  We further assume the
average gas densities are,
\begin{eqnarray*}
{\overline{\rho}}_V &=& 0.1 ~\Omega_b ~\rho_{crit}, \\
{\overline{\rho}}_{NV} &=& 2.0 ~\frac{\Omega_b}{\Omega_m} ~\rho_{crit},
\end{eqnarray*}
where the former suggests the background (unclustered) density of baryons
is a tenth of the mean, while the latter is Eq. \ref{eq-rhogf}.
Thus, $\rho_{NV}/\rho_V \sim 67$ when $\Omega_m = 0.3$.  These values
imply that
\begin{displaymath}
v_{out} \simeq 95 ~\kms.
\end{displaymath}
That is, an outfall velocity of $v_{out} \sim 95$ \kms\ will maintain
a pressure discontinuity consistent with what was presumed here.  This
accounts for the strong density gradient at the boundary between void
and non-void space, and the apparent rapidity of the onset of shock
stripping on clouds; it helps confirm the self-consistency of the
physical picture.

\subsection{The Non-Void Cloud Spatial Distribution}

I now consider what might be learned from the trend of the cumulative
line density of \lya\ absorbers as a function of ${\cal T}_{lim}$.
The trend in the log of the intercept $C$ in Fig. \ref{fig-ewdf}$B$
for the non-void EWDFs (solid line, right-hand side of upper sub-panel
of panel $b$) appears to be well-fit by a line.  I find that the trend
is consistent with a cumulative line density at ${\mathcal W}=63$
m\AA\ of $d{\mathcal N}(\ge{\mathcal T}_{lim})/dz=783 \,{\mathcal
T}_{lim}^{0.394}$.  The figure shows that this relation is accurate
over the range $0.1 \la {\mathcal T}\la 4$.  The slope may be taken to
give information about the radial distribution of clouds about
isolated galaxies, for the steep dependence of tidal field strength
with distance from galaxies essentially ensures that high tidal field
regions are close to a strong concentration of mass.  Consider an
isolated, centrally condensed body.  It follows that the tide varies
as $\sim R^{-3}$ (Eq. 17$b$ of Paper I)\footnote{As noted in \S5.1 of
Paper I, galaxies are taken to extend to the truncation radius $R_t$,
assumed to be $500 ~\h_{75}^{-1}$ \kpc\ for an ${\mathcal L}^*$
galaxy.  In calculating tidal fields, the galaxy mass is summed to
$R_t$, and the tide is calculated as though it were a point mass.
Thus the above method is consistent even within $R_t$.}, so that the
cumulative line density is $d{\mathcal N}(\le R_{lim})/dz \propto
R_{lim}^{-1.18}$, where $R_{lim}$ is the distance at which ${\mathcal
T} = {\mathcal T}_{lim}$.  This implies a differential line density
$d^2{\mathcal N}(R)/dz\, dR \propto R^{-2.18}$.  If the radii of
clouds within the IGM do not vary systematically with tidal field,
then the number density of clouds is proportional to the differential
line density.  This exponent, $-2.18$, is quite close to that derived
for the distribution of satellite galaxies around isolated parent
galaxies found by \citet{McKay:02}; $n \propto R^{-2.1}$ -- a relation
valid in the range $133 \la R \la 670 ~\h_{75}^{-1}$ kpc.  For a case
in which tides are caused by an ${\mathcal L}^*$ galaxy ($v_c \simeq
161$ \kms; \citealt{Tully:00}), with Eqs. 14 and 16 of Paper I, it can
be shown that scalar tidal fields in the range $0.1 \la {\mathcal T}
\la 4$ (inverse Hubble times squared) would be produced at a distance
$730 \la R \la 2500 ~\h_{75}^{-1}$ kpc.  For reference, the high
${\cal T}$-side of the transition zone (${\cal T} \simeq 0.2$ would
occur $\sim 2$ Mpc from an ${\cal L}*$ galaxy, making the half-width
of the transition zone $\sim 0.5$ Mpc.

The lower limit of the tidal field range is enticingly close to the
upper limit of the McKay \etal range.  If these clouds trace mass,
then this apparently quite extended, though sparse, ``halo'' of
non-void clouds about galaxies may represent an extension of the mass
distribution entailed by the flat velocity dispersions around giant
field galaxies that extend out to $R_p \simeq 670 ~\h_{75}^{-1}$ kpc
\citep{McKay:02}.  If this is so, then there is a chance to calibrate
how cloud number densities trace matter density.  This may eventually
lead to an estimate of the void matter density.

This analysis strongly suggests that the observed CHVC population,
thought to be dark matter-held \citep{Blitz:99,Sternberg:02}, and with
an average Local Group barycentric motion of $-100$ \kms\
\citep{Braun:99}, are representatives of sub-galactic halos that have
accreted to the IGM long ago.  They are, I would suggest, the closest
examples of the discrete clouds which constitute the non-void column
density spectrum.

\section{Summary}

A self-consistent picture has been built that uses shock-stripping in
the IGM (non-void space) to transform void clouds into non-void
clouds.  A connection between the relative LF normalizations of the
HVDFs in void and non-void space was used to calculate a
``concentration factor'' that adjusts the CDS of transformed model
void clouds for the convergent concentration of clouds into the more
compact non-void space of the IGM.  Whether the concentration factor
can successfully explain the observed non-void CDS depends on the
cloud velocity and the attendant ram-pressure stripping.  I have shown
that non-void clouds are consistent with shock stripped void clouds
when this correction factor is employed.  When cloud velocities
$v_{cl} \ga 75$ \kms, the shape of the model CDS derived here is in
excellent agreement with the observations when the baryons in shocked
clouds lose their central condensation and become evenly distributed
inside the shear layer.  Mergers in dense regions of non-void space
may increase the number density of ${\cal L}^*$-galaxies, a change
which is accompanied by a lowering of the ``faint-end'' slope
parameter $\alpha$.  This suggests fewer clouds in non-void space (by
up to $\sim 1/3$), lowering $f_{mult}$, and suggesting lower outfall
velocities.  For the lower baryon density initially assumed for this
and previous papers, the implied velocity could be less than 50 \kms,
but does not produce a CDS that fits well with observations.  However,
for the larger baryon density consistent with the WMAP results,
$v_{cl}\approx 100$ \kms\ appears consistent from the standpoint of
the quality of the fit to the observed non-void CDS, and the implied
concentration factor $f_{mult}$.  

Indications from the systematic velocity of CHVCs, and from the
requirements of maintaining a strong pressure gradient between void
space and the non-void environment, appears to require a void
``outfall'' velocity on order $v_{out}\approx 100$ \kms.

\section{Acknowledgments}

I wish thank Christopher F. McKee for many helpful comments and
suggestions.  I thank Hy Spinrad for financial support and for his
wisdom during his stint as my research advisor.  I received financial
support of NSF grant \#AST-0097163 and the UC Berkeley Department of
Astronomy.
%\newpage

%\bibliographystyle{apj} 
%\bibliography{apj-jour,manning}
%%\begin{thebibliography}{46}
%\end{document}

\end{document}